\documentclass{aastex}
\def\be{\begin{equation}}
\def\ee{\end{equation}}
\def\etal{{\it et al. }}

\def\kmsm{km~s$^{-1}$~Mpc$^{-1}$~}

\def\hal{H$\alpha$}
\def\m{$\mu$m}
\def\smalltype{\let\rm=\eightrm \let\bf=\eightbf
  \let\it=\eightit \let\sl=\eightsl \let\mus=\eightmus
  \baselineskip=9.5pt minus .75pt
  \rm}
\begin{document}

\title{Extinction Effects in Spiral Galaxy Rotation Curves}

\author{ Carlos Valotto\altaffilmark{1} and  Riccardo Giovanelli}
\affil{Department of Astronomy, Space Sciences Building, Cornell University,
Ithaca, NY 14853 valotto@astro.cornell.edu, riccardo@astro.cornell.edu}

\altaffiltext{1}{AURA Visiting Fellow. Permanent address: Grupo de Investigaciones en
Astronom\'{\i}a Te\'orica y Experimental, Observatorio Astron\'omico,
Universidad Nacional de C\'ordoba, Laprida 854, C\'ordoba 5000, Argentina.}

\begin{abstract}
Observations show that the slope of the inner part of the H$\alpha$/[NII]
rotation curves of disk galaxies is depressed by extinction: at fixed
luminosity, the observed slope is in fact seen to depend on the disk
inclination to the line of sight. Using a simple extinction model, we
are able to reproduce well the observed trends. The model assumes an
exponential distribution, both in the radial and $z$ directions, identical
for star--forming regions and dust. Fits to the data are optimized by
varying the scale height and scale length of absorbers, as well as the
disk's central optical depth $\tau_\circ$, as seen face--on. The
observations indicate that disk central opacity increases with total
luminosity. Model fits for the most luminous galaxies (brighter than
$M-5\log h=-21.4$ in the I band) yield $\tau_\circ \simeq 3.5^{+4.0}_{-2.0}$,
near $\lambda=0.66$ $\mu$m. The large uncertainty on the inferred value
of $\tau_\circ$ is due to the poorly known characteristics of the
distribution of absorbers in the disk, as well as to the likelihood of
strong departures from an exponential radial distribution near the
galaxy centers.

\end{abstract}


\section{Introduction}

Scattering and absorption by interstellar dust depresses the observed flux
of galaxy disks at optical and infrared wavelengths. While the magnitude of
the effect has been disputed, there is no doubt that the effect is strongly
dependent on the inclination of a disk to the line of sight. A number of
techniques have been used in observationally deriving extinction laws (see
for example the volume edited by Davies and Burstein 1995, the review by
Calzetti 2001 and Masters, Giovanelli \& Haynes 2003 for details). 
Goad
and Roberts (1981) first discussed a
kinematical technique which we briefly describe as follows.

Let $V(r)$ be the rotational velocity of an axially symmetric disk at the
distance $r$ from its center and $(x,y)$ be a set of Cartesian coordinates
in the plane of the disk. If the disk is thin and its symmetry axis
is inclined by an angle $i$ to the line of sight, the observed component
of velocity which intercepts the disk at $(x,y)$
is
% EQN (1)
\be
V_\parallel = V(r) {x\over \sqrt{x^2+y^2}} \sin i + V_{nc},
\ee
where $V_{nc}$ accounts for non--circular motions and the $x$ axis is oriented
along the disk's apparent major axis. The observed rotation curve, as derived
for example from a long--slit \hal ~spectrum positioned along the major axis,
will depart from the prescription given by Eqn. 1 as due to seeing,
instrumental resolution, averaging across the slit  width, the finite thickness
of the disk and extinction occurring within the disk itself. As a realistically
thick disk approaches the edge--on perspective, lines--of--sight along the major
axis sample regions of increasingly broad range in $y$, yielding a velocity
distribution with a peak velocity contributed by parcels of gas asymmetrically
distributed with respect to $y=0$: if extinction is important, only foreground
parts of the disk contribute to the emission and the factor $(x/ \sqrt{x^2+y^2})<1$
depresses the velocity distribution observed at $r=x$.
Goad \& Roberts noted how this tapering effect may, in opaque edge--on
disks, yield observed rotation curves resembling solid--body behavior,
independently of the true shape of $V(r)$. Bosma \etal (1992) tested this
technique on two edge--on systems, NGC 100 and NGC 891. They compared HI
synthesis and \hal ~observations and concluded that the disk of NGC 100 is
transparent, while in the case of NGC 891 they could not exclude the possibility
of extinction in the inner parts of the disk. Their conclusions were however
strongly influenced by the limited spatial resolution of the HI data. Prada
\etal (1994) compared long--slit spectra of NGC 2146 in the optical and near
IR and reported evidence for extinction in the inner parts of the galaxy.
More recently, Giovanelli \& Haynes (2002, hereafter GH02) applied the same technique
in a statistical fashion, detecting clear evidence for the effect of
extinction at 0.66 \m. Using a sample of more than 2000 \hal ~rotation
curves, they found that the opacity in the inner disks of spiral galaxies
is luminosity dependent, in a manner previously found by purely photometric
means (Giovanelli \etal 1995; Tully \etal 1998).
The observational data of GH02 are corroborated by the detailed radiative
transfer calculations of Baes et al.(2004), which also obtain dramatic
increases in the apparent inner scale length of rotations curves
as disks approach the edge--on perspective.

In this report, we use a simple model for the dust gas distribution in spiral
disks, in order to reproduce the findings of GH02 and
in so doing to obtain quantitative inferences on the characteristics of the
disk opacity. In Section 2 we summarize the observational results that are
to be modelled. In Section 3 we describe the disk model to be fitted to the
data, while the best fit parameters are discussed in Section 4. In Section 5
we present our conclusions. Throughout this
work, distance dependent quantities are scaled according to a Hubble parameter
$H_\circ=100 h$ \kmsm.

\section {Rotation Curve Shape and Disk Inclination}
%SECTION 2

GH02 fitted \hal/[NII] rotation curves of spiral galaxies
with the parametric model
\be
V_{pe}=V_\circ (1-e^{-r/ r_{pe}})(1+\alpha r/ r_{pe})
\label{vpe}
\ee
where $V_\circ$ scales the amplitude of the rotation curve, $r_{pe}$ yields a
scale length for the inner steep rise, and $\alpha$ corresponds to the slope
at large $r$. In their Table  1, for different luminosity classes as measured
in the I--band they list mean
values of $\alpha$, $h r_{pe}$ and of the ratio between $hr_{pe}$ and the scale
length of the disk light for relatively face--on systems. In their Figure 1,
for each luminosity class the dependence of $h r_{pe}$ on disk inclination $i$
is shown. For a given luminosity class, with increasing $i$, $<h r_{pe}>$ remains
at first constant; as $i$ approaches $70^\circ$, $<h r_{pe}>$ starts increasing,
the rate of increase being larger for more luminous systems. In the latter,
$<h r_{pe}>$ at high inclination has values more than double the values observed
at low inclination.  The inclination dependence was interpreted as due to the
increase in the opacity of spiral disks with inclination, while the luminosity dependence was attributed to a combination of increased
disk size and higher interstellar medium metallicities in the more luminous systems.
By contrast, the outer slope $<\alpha>$ is unaffected by disk inclination.

\section{Disk Extinction Model}
%SECTION 3

As done by other authors who have studied disk extinction models (e.g
(Byun \etal ~1994; Xilouris \etal ~1999; Misiriotis \etal ~2000), we
assume that the emissivity per unit volume is produced in the disk by
a distribution of isotropic emitters which is exponential both in the
radial direction $r$ in the plane of the disk, and in the direction $z$
perpendicular to it, so that
\be
j(r,z)=j_\circ ~ e^{-r/r_d} ~ e^{-|z|/z_d}  \label{eqj}
\ee
where $r_d$ and $z_d$ are respectively the scale length and the scale height,
and we ignore the effects of clumpiness and azimuthal variations.
We will need to distinguish between scale lengths and scale heights of
stars, dust and \hal--emitting regions, respectively referred to by
subscripts ``*'', ``d'' and ``$\alpha$''.
We will assume that the dust distribution in the disk is exponential
both in $r$ and $z$, with scale length and scale height of the dust
as those of the \hal ~light at $\lambda=0.66$ $\mu$m. That assumption
greatly simplifies the analysis; it is justified  since the light
distribution in our case is that of the \hal ~produced in star forming
regions, which is likely to mimick that of dust clouds. It is
however a simplistic assumption, which cannot fully reproduce the
more complex characteristics of the distribution of star--forming
regions in galaxies. We use it here because of its simplicity, and
discuss later some of its possible limitations. The absorption
coefficient per unit volume for an exponential distribution can thus
be written as
\be
\kappa(r,z)=\kappa_\circ ~ e^{-r/r_{d,d}} ~ e^{-|z|/z_{d,d}}   \label{eqk}
\ee
The optical depth at the disk center, when seen face--on, is then
$\tau_\circ = 2\kappa_\circ z_{d,d}$, while the optical depth as a function
of the spatial coordinate along the line of sight $s$ is
$\tau(s) = \int^{\infty}_s\kappa(s')ds'$. For a randomly oriented disk,
the optical depth to any parcel of emitting gas at location $s$ is fully
specified once $\tau_\circ$, $r_{d,d}$, $z_{d,d}$ and $i$ are given. The
disk's intrinsic axial ratio for each component, $q_\circ = z_d/r_d$, is
related to the apparent ratio between minor and major radii $b/a$ and the inclination to the line of sight via
\be
\cos^2 i = [(b/a)^2-q_\circ^2] / (1 - q_\circ^2),   \label{eqcosi}
\ee
which is needed to relate the model to the observations, for
GH02 show the inclination dependence of the rotation curve shape by
plotting $h r_{pe}$ vs. $\log_{10} (a/b)$. GH02 derived axial ratios
from I--band surface photometry, a band for which the light originates
in late disk stars; the $q_{\circ,*}$ of that light is very likely to be significantly larger than that of the dust and of the star--forming regions.
In Giovanelli \etal (1994,1995) they derived a value of $q_{\circ,*}$
of 0.13 for the I--band. The largest data sets of \hal ~surface
photometry are those of Ryder \& Dopita (1994) and Koopmann, Kenney
\& Young (2001). Ryder and Dopita have compared stellar and \hal~
scale lengths, showing the latter to be significantly larger than
the former; from their figures 4 and 5, we infer
$r_{d,\alpha}=(1.9\pm0.4)r_{d,*}$, where the stellar scale length
corresponds to the I--band.

The contribution to the observed flux along a given line of sight by a parcel
of gas at location $s$ is proportional to $j(s)e^{-\tau(s)}$. The \hal ~line
profile along a given line of sight through the disk will be
$\propto \int_0^\infty\,j(s)e^{-\tau(s)}ds$, where $s$ is mapped onto the line
of sight velocity, given a rotation curve $V(r)$.
In analogy with the measurement practice by which the rotation curve of a disk
is extracted from a slit positioned along the disk major axis, in our model we
assign to each line of sight the velocity corresponding to the peak of the
simulated \hal ~line profile.

We use the model as follows:
\begin{itemize}
\item Given a disk galaxy luminosity class, we obtain an ``unextincted'' rotation
curve $V(r)$ of the type described by Eqn. \ref{vpe}, using the average values
of the structural parameters as given in Table 1 of GH02.
\item We select a set of disk parameters $r_{d,d}$, $q_{\circ,d}$
--- the same as for the \hal~ light --- and $\tau_\circ$ and
build an ``observed'' rotation curve, that takes into account the effects of
extinction as produced by the adopted disk model. We then measure $h r_{pe}$ in
the thus modified rotation curve.
%CAMBIO RESPECTO AL ANTERIOR
\item By progressively changing the inclination angle of the disk, we obtain
a curve that describes the change of $h r_{pe}$ with $\cos(i)$,
which we compare with the observed data as shown in Figure 1 of GH02.
In order to show the dependence with $\cos(i)$ in the observed data,
we use the Eqn. \ref{eqcosi}.
\item The comparison between model and data is optimized via the minimization of
\be
\chi^2=\sum_{i=1}^N \left[\frac{r_{pe,i}-r_{pe}
[\cos(i)_i;\tau_\circ,q_{\circ,d},r_{d,d}]}
{\sigma_i}\right]^2  \label{chi}
\ee
where $[r_{pe,i},\cos(i)_i]$ is the i$th$ point of the data to be fitted,
and $\sigma_i$ is its associated uncertainty. The procedure yields a best
fit of unconstrained values for $r_{d,d}$, $q_{\circ,d}$ and $\tau_\circ$,
or best fit values of any of those parameters, when the others
are constrained. Since GH02 express the disk inclination in terms of the
axial ratio of the starlight in the I--band, we use Eqn. (5) with $q_\circ=q_{\circ,*}=0.13$ to obtain $\cos (i)$.
\end{itemize}

\section{Model Fitting Results}
%SECTION 4

The GH02 analysis of rotation curves separates the available sample in
six luminosity classes. The statistically more significant trend of
$h\,r_{pe}$ with inclination takes place for the two brightest
luminosity classes: Class I includes galaxies with $M-5\log h<-22$
and Class II includes galaxies with $-22.0<M-5\log h<-21.4$. The
trends in $h\,r_{pe}$ with inclination are shown by the data points
in Figure 3. The average stellar scale lengths in the I band are
respectively $h\,r_{d,*}=4.3$ kpc and $h\,r_{d,*}=3.2$ kpc, for
Classes I and II, as given by Haynes \etal ~1999. Here, we restrict
our analysis to those two classes. For the less luminous classes,
extinction is small, the inclination dependence of $h r_{pe}$ is marginal
or nonexistent and the value of $\tau_\circ$ that would be derived
is unreliable.

Figures 1 and 2 show simulated rotation curves, as obtained using our
disk extinction model, Eqn. \ref{vpe} ~and the parameters listed in
Table 1 of GH02, for the two luminosity classes of interest.
In Figure 1, the ``unextincted'' rotation curve for Lumnosity Class
I, normalized by a factor $V_\circ \sin i$  is shown as a thick line
in each of the two panels. In panel (a), the effect of changing the
disk inclination is shown,
with thin lines exhibiting rotation curves for values of $i$ between
$70^\circ$ and $86^\circ$, $\Delta i =4^\circ$,
at a fixed value of $\tau_\circ=3.0$, for $h\,r_{d,d}=8.6$ kpc and
$q_{\circ,d}=0.05$, plotted versus the radial distance in units of
the stellar scale length $r_{d,*}$.
In panel (b), for the same values of $h\,r_{d,d}$ and $q_{\circ,d}$,
the effect of varying the central disk
optical depth is shown, whereby for a fixed inclination of
$i=86^\circ$, the values for $\tau_\circ=0.5, 1.0, 2.0, 5.0, 10.0$,
are shown in the corresponding curve.
Figure 2 displays the analogous model curves for Luminosity Class II,
for which we have assumed $h\,r_{d,d}=6.4$ kpc and $q_{\circ,d}=0.05$.
In all panels, the intersection of the horizontal
dashed line and the rotation curve identifies the corresponding value of
$r_{pe}$ for that curve. Note how even for disks of moderate opacity,
the value of the rotational velocity within a few scale lengths from the
center can be drastically altered by extinction, in the case of highly
inclined and luminous systems.

Figure 3 shows the GH02 data on the variation of $h r_{pe}$ with inclination,
and the corresponding model trends as derived from our disk extinction model,
for varying values of $\tau_\circ$ (1 to 10, $\Delta \tau_\circ=1$).
In each case, $q_{\circ,d}$ and $h r_{d,d}$ are the same values used in
Figures 1 and 2. The models plotted with a thick line correspond to
$\tau_\circ=5.0$.

In order to verify the variance in the best fit value $\tau_\circ$
produced by different values of the assumed $q_{\circ,d}$,
Figure 4 shows the data for the two luminosity classes and, separately,
simulated $h r_{pe}$ vs. $\cos (i)$ relations for $\tau_\circ$ between
1 and 10 and for $q_{\circ,d}=0.03$ (panels a, d), 0.05 (panels b, e)
and 0.07 (panels c, f). Similarly, Figure 5 shows the variance
introduced by different assumptions on the dust disk scale length
$h r_{d,d}$: for $q_{\circ,d} =0.05$, the model curves were computed
for $h r_{d,d}=7.0$ kpc (panel a), 8.0 kpc (panel b), and 9.0 kpc
(panel c) for Luminosity Class I, and 5.5 kpc (panel d), 6.5 kpc
(panel e) and 7.5 kpc (panel f) for Luminosity Class II.
For given $\tau_\circ$, disk inclination and $r_{d,d}$, a higher
value of $q_{\circ,d}$ implies that extinction along a given
path length involves a wider interval of disk radii, thus resulting
in a more ``smeared'' rotation curve (i.e. longer $h\,r_{pe}$).
Similarly, for given $\tau_\circ$, disk inclination and $q_{\circ,d}$,
a longer dust scale length also implies that extinction along a
given path length involves a wider interval of disk radii, thus
a longer $h\,r_{pe}$. Both effects are seen in the simulations.

Unconstrained fits of the type described by Equation (6) can yield
simultaneously values for $\tau_\circ$, $q_{\circ,d}$ and $r_{d,d}$.
However, values of $q_{\circ,d}$ and $r_{d,d}$ are not independent:
lengthening the disk will obviously reduce its axial ratio. We can thus
maintain $<r_{d,d}>$ and/or $<q_{\circ,d}>$ fixed and obtained constrained
fit values for $<\tau_\circ>$ for each Luminosity Class. What is a
reasonable range of values for $<q_{\circ,d}>$ and $<r_{d,d}>$ ?

While photometric measurements provide a fair assessment of the mean
and variance in the value of $h r_{d,*}$ for a given luminosity class
(the scatter about the observed mean values is approximately 0.08 in
$\log r_{d,*}$), the
distribution in disk intrinsic axial ratios is quite uncertain, for
$q_{\circ,*}$ is a difficult quantity to measure and even its mean value
for a given type of galaxy relies on poorly constrained statistical
inferences from the observed distribution of axial ratios and from a
few observations of edge--on systems. The mean value of
$q_{\circ,*} = 0.13$ was adopted by Giovanelli \etal(1994) for the
I band. The extreme disk population of star forming regions and dust
should be significantly thinner than that of the solar--type disk stars
contributing most of the light at I band. Ryder \& Dopita (1994)
give a comparison between $r_{d,*}$ and $r_{d,\alpha}$, which we
assumed to be the same as $r_{d,d}$. No equally accessible comparison
between $q_{\circ,*}$ and $q_{\circ,d}$ is available. In Table 1,
we give estimates of $\tau_\circ$ obtained assuming the tabulated
values for $r_{d,d}$ and $q_{\circ,d}$, based on the Ryder \&
Dopita (1994) data and heuristic assumptions on the dust layer
intrinsic axial ratio. The error assigned to this determination
is based on a combination of (a) the standard error propagation
analysis derived from the impact of the GH02 data quality, (b)
uncertainty in the conversion of $r_{d,*}$ to $r_{d,d}$, estimated
to be 20\% on the basis of the Ryder \& Dopita data, and (c)
uncertainty in the value of $q_{\circ,d}$, assumed to hover between
0.035 and 0.06 for both Luminosity Classes. The largest source of
uncertainty is clearly that deriving from $q_{\circ,d}$. An additional
source of uncertainty is introduced by the often--seen departures from
an exponential form, in the radial distribution of \hal ~especially
near the centers of galaxies: a cursory inspection of the Ryder \&
Dopita and the Koopmann \etal ~samples indicates that about half
of galaxies exhibit a cusp in the \hal ~distribution, within the
inner 1--2 kpc. This region coincides with that within which $r_{pe}$
is measured. Thus, values of $\tau_\circ$ inferred with the assumption
that the radial distribution of \hal, and thus of dust, is exponential
throughout the disk may underestimate the opacity of the inner regions
of the disk and overestimate it in the outer regions. While it
is clear that extinction plays a role in affecting the shape of the
inner parts of \hal~ rotation curves of bright spirals, the technique
outlined in this paper for its measurement requires higher quality
estimates of the distribution of absorbers than currently available,
for $\tau_\circ$ to be pinned down to better than within a factor of
2 or so.

\begin{deluxetable}{cccll}
\tablecaption{Inferred Disk Central Optical Depth $\tau_\circ$}
\tablewidth{0pt}
\tablehead{\colhead{Lum Class} & $h\,r_{d,*}$ (kpc) & $h\,r_{d,d}$ (kpc) &
\colhead{$q_{\circ,d}$} & \colhead{$\tau_\circ$}}
\startdata
I   & 4.3 & $8.2\pm 1.7$ & $0.045\pm0.015$ &$2.5^{+2.5}_{-1.5}$  \\
II  & 3.2 & $6.4\pm 1.2$ & $0.050\pm0.015$ &$3.5^{+4.0}_{-2.0}$  \\
\enddata
\end{deluxetable}

\section{Conclusions}

A simple model, whereby dust and HII regions
are distributed exponentially in both the radial and $z$ directions,
with equal scale height and scale length, can reproduce the observed
variation in the inner slope of H$\alpha$/[NII] rotation curves of
disk galaxies. The effect is due to internal extinction. For reasonable
values of the ratio between scale height and scale length, the model
can constrain central disk opacities for the most luminous
galaxies, in which the effect of extinction is the strongest.
However, pinning down the exact value of central opticall depth of the
disk at 0.66 $\mu$m is difficult, due to uncertainty in the form of
the distribution of absorbers. If \hal ~and the absorbers have
scale length about 1.9 times that of the I band starlight and
scale height to scale length ratio of $\sim 0.05$, we infer
central optical depth of disk, in the face--on perspective of
$\tau_\circ\simeq 3.5$, for galaxies brighter than $M-5\log h=-21.4$
in the I  band. However, the uncertainty in the radial
and $z$--distributions of absorbers constrains the inferred values
of $\tau_\circ$ to an uncertainty no better than a factor of 2.
For less luminous spiral galaxies, extinction becomes progressively
less important, and the model cannot usefully constrain the disk opacity.

\acknowledgments

Support for this work was provided by the National Science Foundation
through grant number GF-1003-00 from the Association of Universities for
Research in Astronomy, Inc., under NSF Coorporative Agreement No. AST-9613615.

{}

\begin{figure}
\plotone{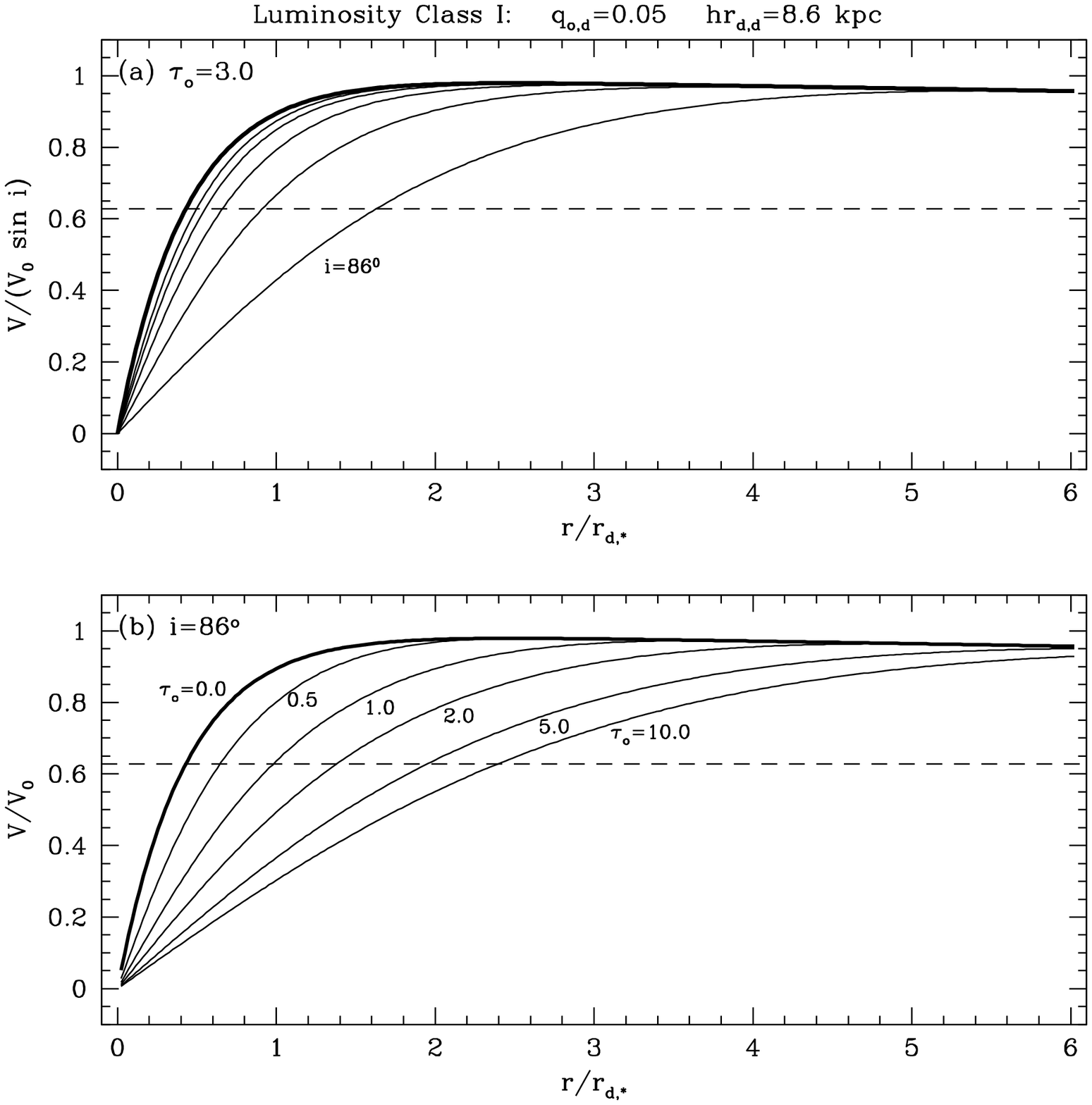}
\caption{Simulated rotation curves for luminosity class I.
The solid line corresponds to the enextincted disk. Panel
(a) has rotation curves for fixed $\tau_\circ = 3$ and varying
disk inclination, from $i=70^\circ$ to $i=86^\circ$. In panel
(b), rotation curves are for fixed $i=86^\circ$ and varying
$\tau_\circ$, as indicated. The intersection of the horizontal,
dashed line and each rotation curve occurs near the radius
$h r_{pe}$.}
\end{figure}

\begin{figure}
\plotone{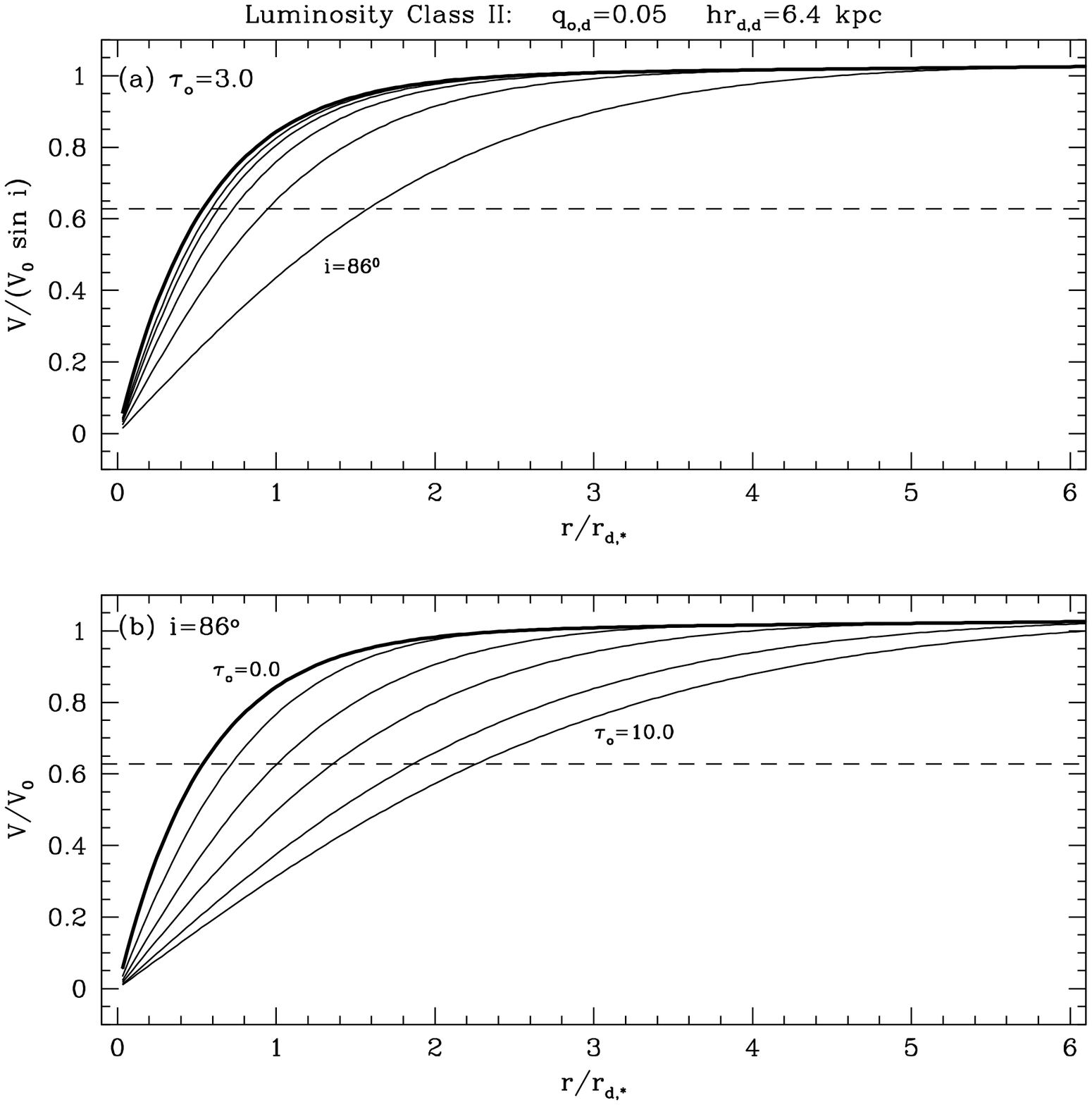}
\caption{Simulated rotation curves for luminosity class II.
The solid line corresponds to the enextincted disk. Panel
(a) has rotation curves for fixed $\tau_\circ = 3$ and varying
disk inclination, from $i=70^\circ$ to $i=86^\circ$. In panel
(b), rotation curves are for fixed $i=86^\circ$ and varying
$\tau_\circ$, as indicated in figure 1. The intersection of the horizontal,
dashed line and each rotation curve occurs near the radius
$h r_{pe}$.}
\end{figure}

\begin{figure}
\plotone{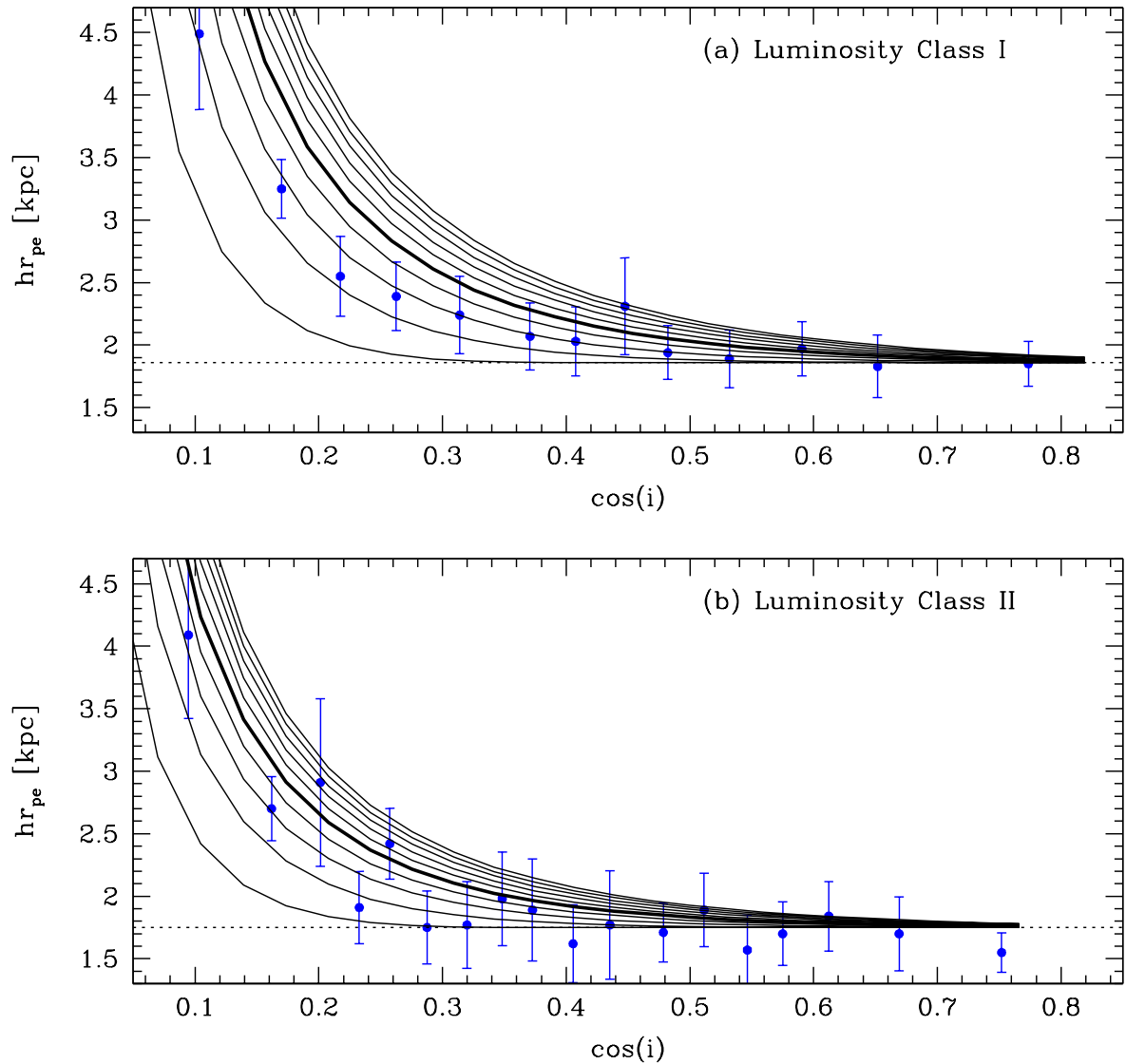}
\caption{GH02 data on the variation of $hr_{pe}$ with inclination
for luminosity classes I and II (data points) and simulated inclination
dependences as obtained by our model, varying values of $\tau_\circ$
between 1 and 10 (bottom to top solid lines). The thick line corresponds
to $\tau_\circ=5$ in each panel.}
\end{figure}

\begin{figure}
\plotone{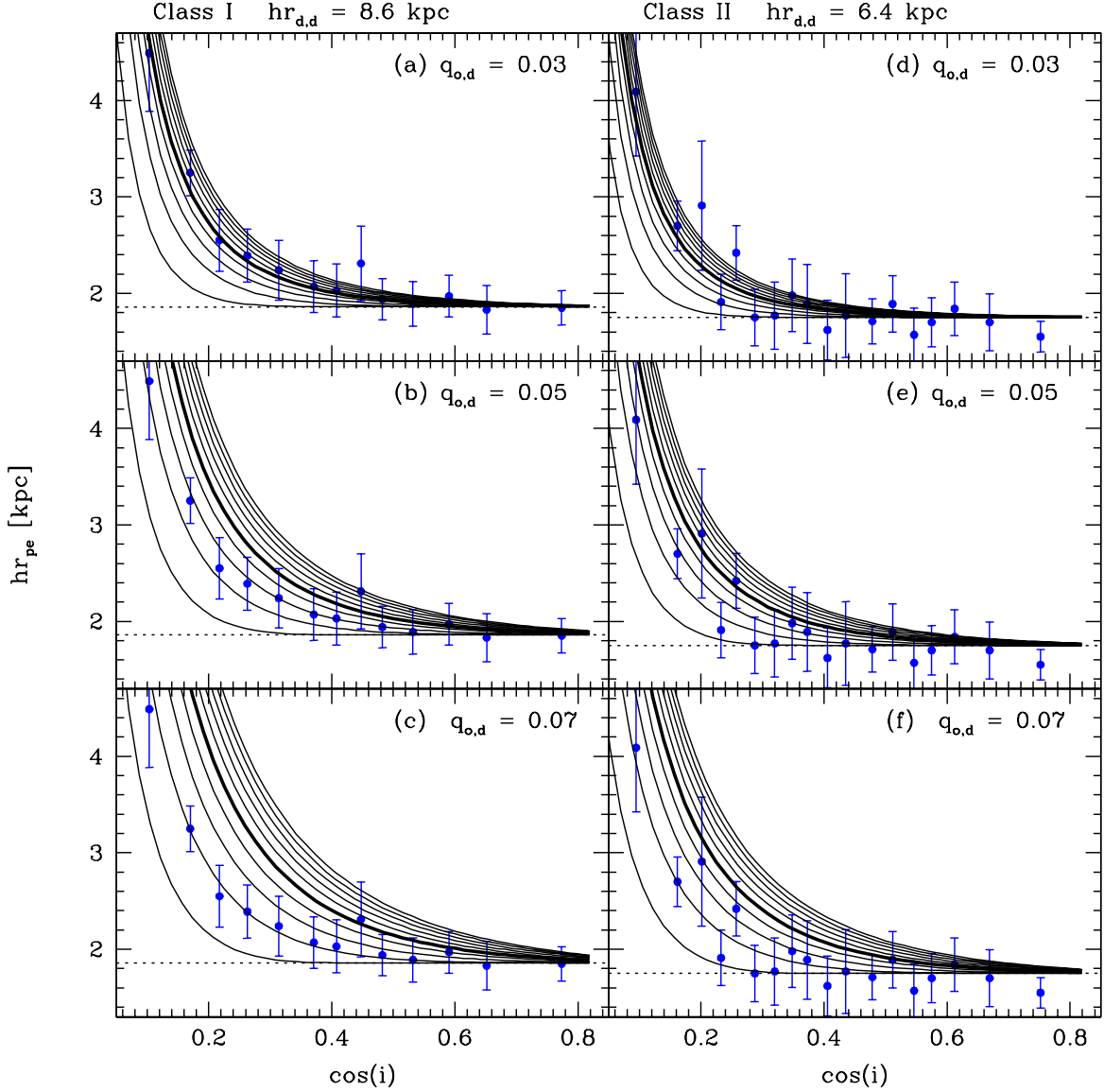}
\caption{$hr_{pe}$ vs. $cos(i)$ for Luminosity Class I
(Luminosity Class II)
varying $\tau_\circ$ between 1 and 10 (bottom to top solid lines),
for
panel a(d) $q_{\circ,d} = 0.03$,
panel b(e) $q_{\circ,d} = 0.05$ and
panel c(f) $q_{\circ,d} = 0.07$.
The thick line corresponds to $\tau_\circ=5$ in each panel.
}
\end{figure}

\begin{figure}
\plotone{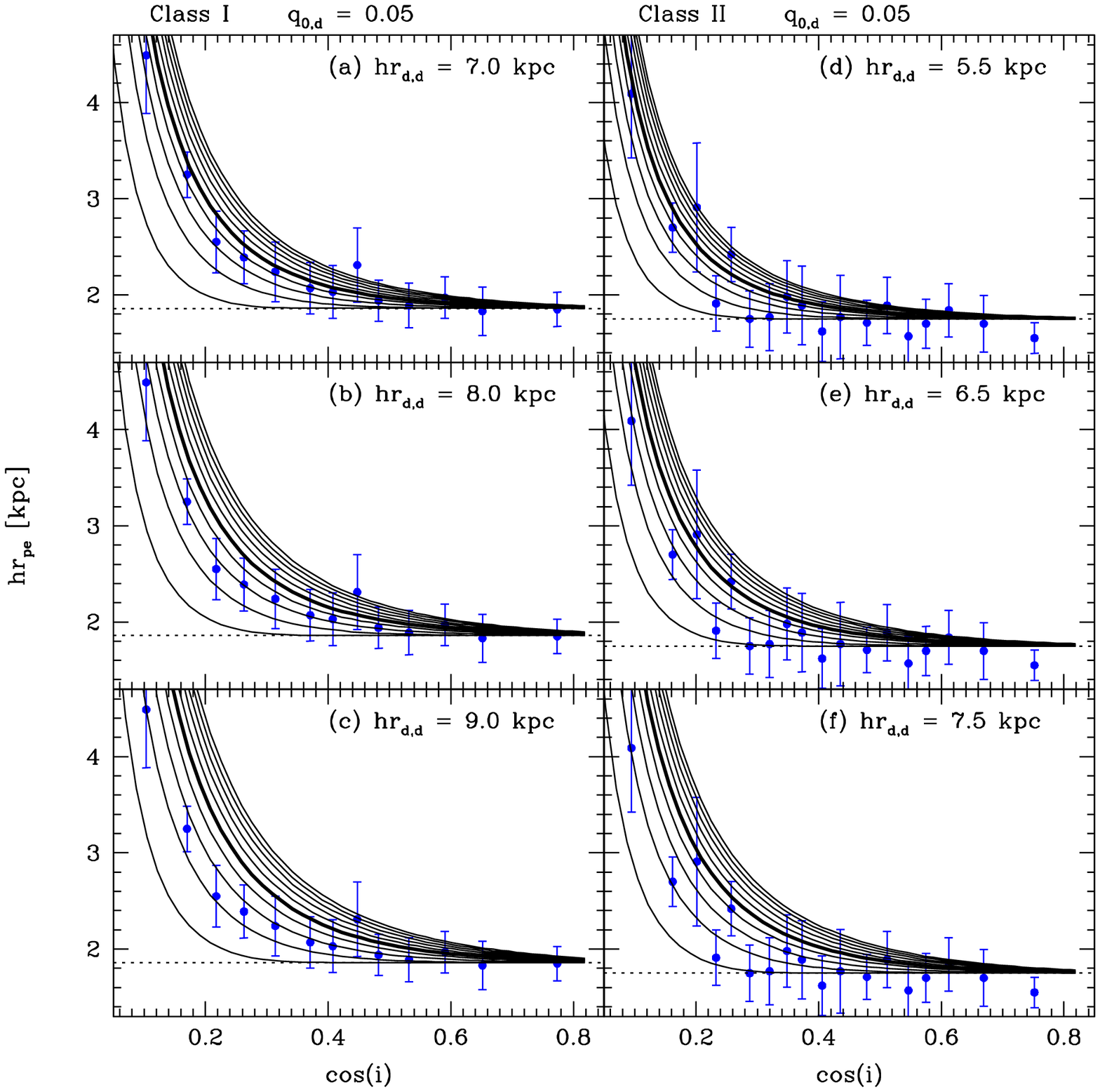}
\caption{$hr_{pe}$ vs. $\cos(i)$ for Luminosity Class I varying
$\tau_\circ$ between 1 and 10 (bottom to top solid lines),
for
(a) $h r_{d,d} = 7.0$ kpc,
(b) $h r_{d,d} = 8.0$ kpc and
(c) $h r_{d,d} = 9.0$ kpc,
and Luminosity Class II for
(d) $h r_{d,d} = 5.5$ kpc,
(e) $h r_{d,d} = 6.5$ kpc and
(f) $h r_{d,d} = 7.5$ kpc.
The thick line corresponds to $\tau_\circ=5$ in each panel.
}
\end{figure}


\begin{thebibliography}{}

\bibitem[Bos 1992]{}
Bosma, A, Byun, Y., Freeman, K.C. \& Athanassoula, E. 1992, \apj ~400, L21

\bibitem[]{} Baes, M., Davies, J.I., Dejonghe, H., Sabatini, S.,
Roberts, S., Evans, R., Linder, S.M., Smith, R. \& de Blok, W.J.G. 
2003, \mnras, 343, 1081

\bibitem[]{}
Calzetti, D. 2001, \pasp ~113, 1449

\bibitem[Byun, Freeman \& Kylafis(1994)]{byu94} Byun, Y.Y., Freeman, K.C. \&
Kylafis, N.D., 1994, \apj ~432, 11.

\bibitem[]{}
Davies, J.I. \& Burstein, D. 1995, editors: {\it The Opacity of Spiral
Disks}, NATO ASI Series vol. 469, Kluwer:Dordrecht

\bibitem[Giovanelli et al.(1994)]{gio94} Giovanelli, R. Haynes, M.,
Salzer, J., Wegner, G., Da Costa, L., \& Freudling, W., 1994, \aj~ 107, 2036

\bibitem[Giovanelli et al.(1995)]{gio95} Giovanelli, R. Haynes, M.,
Salzer, J., Wegner, G., Da Costa, L., \& Freudling, W., 1995, \aj~ 110, 1059

\bibitem[Giovanelli \& Haynes (2002)]{gio02} Giovanelli, R. \&
Haynes, M. 2002, \apj~ 571, L107  (GH02)

\bibitem[Goa 1981]{}
Goad, J. \& Roberts, M.S. 1981, \apj ~250, 79

\bibitem[]{} Haynes, M., Giovanelli, R.,
Salzer, J., Wegner, G., \& Freudling, Da Costa, L., W., Herter, T.,
Vogt, N.P. 1999, \aj~ 117, 1668

\bibitem[]{} Koopmann, R.A., Kenney, J.D.P. \& Young, J. 2001, \apjs ~135, 125

\bibitem[]{} Masters, K.L., Giovanelli, R. \& Haynes, M.P. 2003,
\aj~ 126, 158

\bibitem[Misiriotis et al.(2000)]{mis00} Misiriotis, A., Kylafis, N.D.,
Papamastorakis, J., \& Xilouris, E.M., 2000, \aap~ 353, 117

\bibitem[]{} Prada, F., Beckman, J.E., McKeith, C.D., Castles, J. \& Greve, A.
1994, \apj~ 423, 35 

\bibitem[]{} Ryder, S.D. \& Dopita, M.A. 1994, \apj ~430, 142

\bibitem[Tul 1988]{tul88}
Tully, R.B., Pierce, M.J., Huang, J.S., Saunders, W., Verheijen, M. \&
Witchalls, P. 1998, \aj ~115, 2264

\bibitem[Xilouris et al.(1999)]{xil99}Xilouris, E.M., Byun, Y.I., Kylafis, N.D.,
Paleologou, E.V. \& Papamastorakis, J., 1999, \aap~ 344, 868

\end{thebibliography}
\end{document}